\documentclass[12pt]{iopart}
\usepackage{float}
\usepackage{graphicx}
\usepackage[UKenglish]{babel}
\usepackage{url}
\usepackage[numbers,sort&compress]{natbib}

\begin{document}
	\title{Quantum key distribution with displaced thermal states}
	\author{A Walton$^{1,2}$, A Ghesqui\`ere$^1$ and B Varcoe$^1$}
	\address{$^1$ School of Physics and Astronomy, University of Leeds, Leeds, LS2 9JT, United Kingdom}
	\address{$^2$ Email: pyaw@leeds.ac.uk}
	\begin{abstract}
 
Secret key exchange relies on the creation of correlated signals, serving as the raw resource for secure communication. Thermal states, exhibit Hanbury Brown and Twiss correlations, which offer a promising avenue for generating such signals.
In this paper, we present an experimental implementation of a central broadcast thermal state quantum key distribution (QKD) protocol in the microwave region. Our objective is to showcase a straightforward method of QKD utilizing readily available broadcasting equipment.
Unlike conventional approaches to thermal state QKD, we leverage displaced thermal states. These states enable us to share the output of a thermal source among Alice, Bob, and Eve via both waveguide channels and free space. Through measurement and conversion into bit strings, our protocol produces key-ready bit strings without the need for specialized equipment. By harnessing the inherent noise in thermal broadcasts, our setup facilitates the recovery of distinct bit strings by all parties involved. 
	\end{abstract}
	\maketitle

\section{Introduction}
Traditionally, optical communication has been the cornerstone of quantum key distribution (QKD), relying on proven technologies like lasers and optical fibers capable of transmitting information over long distances and achieving reasonable bit rates. However, optical frequencies can often be unwieldy and impractical for short-range applications, such as communication between mobile devices, medical implants, or electronic car keys and locks. This challenge primarily arises from the `last mile problem,' where difficulties in the alignment (the pointing accuracy) of 
a narrow laser source hinders efficient communication. An additional issue in the optical region involves the regular need for line-of-sight, or a fibre connection between the relevant parties, which may be impractical or expensive to implement.
Recognising the significant gap between optical QKD and the practical demands of short-distance communication, a recent shift in focus has been directed towards exploring microwave QKD\cite{Casariego2023,Candia2021}. This is true in general for communication systems where microwave and radio links are preferred methods for short-range communication, while high-speed optical fibers excel in linking hubs\cite{Casariego2023}. This is true also at a smaller scale where optical fibres come to a house, and WiFi distributes the connections to mobile devices.

Quantum communication has witnessed remarkable strides in the last few years starting with the launch of a quantum satellite in 2016[1] that has recently been used to transmit quantum key over 1200 km\cite{Yin2020}. This marked a significant step towards realizing global scale quantum key distribution (QKD). However, the last couple of years have also witnessed a number of other efforts to explore QKD in a range of frequency regimes \cite{Casariego2023,Candia2021,Zhang2022,Nadlinger2022,Ray2024,Belenchia2022,Li2022,Mountogiannakis2022,Wang2021,Zhong2021,Walton2021}. 

Recently is has been recognised that a thermal state is a useful resource for QKD
\cite{Zhong2021,Walton2021,Newton2020,Ghesquiere2021,Newton2019,Qi2018,Qi2020,Xu2021,Wu2019,Wu2020} and this project aims to develop a practical system to enable thermal state QKD. 
Thermal radiation is useful because it exhibits `bunching,' resulting in high levels of noise correlation. This correlation gives rise to quantum discord \cite{Ragy2013}, which has been long recognised as a necessary condition for QKD \cite{Pirandola2014}.
However, one of the problems with a thermal state is the transmission of the state to the receiver. 
By their very nature, thermal states lack coherence, and even though the thermal photon numbers in the microwave are large, on its own a microwave thermal state is not capable of maintaining a significant signal strength for transmission.

The primary goal of this research is to experimentally demonstrate the use of thermal states as a resource for QKD. 
In this paper, we introduce a new approach where we use displaced thermal states. 
Unlike conventional thermal states, which are centered at zero amplitude and have uncertain phase, displaced thermal states acquire the phase of the displacing coherent state while retaining the noise characteristics of the thermal state. 
The security for continuous variable QKD comes from vacuum noise fluctuations. The equivalent noise in a thermal state can be orders of magnitude larger than that of an equivalent coherent state making displaced thermal states an attractive target for secure communication.

Our goal is to explore how this novel thermal resource can be harnessed for secure communication and by understanding the basics of displaced thermal states and their practical implications, we hope to contribute to making communication more accessible and reliable.
The benefit to using a displaced thermal state is that the state acquires a phase. This has the advantage of allowing us to use of off-the-shelf radio equipment. In the case of long range transmissions, free space transmissions could also allow us to remove the need for a waveguide, which is costly to put in place.

To create correlated measurements for key distillation, we use photon bunching in microwave states emitted by a thermal source located separately from Alice and Bob's detectors \cite{Photon_Bunching}. This effect, causing variations in intensity, persists through a beam splitter. 
Heterodyne detection on the output beams generates the required correlated measurement strings \cite{HBT}. However, the inherent randomness in the beam intensities leads to imperfect correlations, ensuring secrecy between Alice and Bob's measurements \cite{Newton2019,Newton2020,Ghesquiere2021,Walton2021}.
For practical security reasons this source is considered to be under the control of Alice. However, unlike CVQKD, the random modulation is introduced by thermal fluctuations, for this reason it has acquired the moniker `Passive state CVQKD'\cite{Qi2018}. 

From Bob's (and Eve's) perspective, the outputs of a thermal source resemble those of sources used in a Gaussian Modulated Coherent State (GMCS) protocol \cite{GMCS_QKD}. This protocol involves coherent states drawn from a Gaussian distribution and broadcast by Alice. The statistical similarity allows us to apply the same security proofs valid for GMCS QKD to the thermal state protocol, including allowance for finite key effects \cite{Jain2022,Newton2019}, and composable security \cite{Jain2022,Mountogiannakis2021, Leverrier2015}.

Previous studies have shown that correlations from the Hanbury Brown and Twiss effect in thermal states can produce bit strings suitable for key distillation \cite{Newton2019,Newton2020,Ghesquiere2021,Walton2021}. Simulations demonstrate the protocol's resilience against beam splitter attacks from Eve \cite{Walton2021}, and successful simulations of a displaced thermal state protocol suggest feasibility for experimental implementation using current communications equipment, as these states are already employed in devices using Phase-Shift Keying (PSK) in modern signal processing \cite{Displaced_Thermal_State_QKD}.

In our experimental setup, we conduct a thermal state QKD protocol using a pair of radio transceivers. We will now outline the apparatus that we used in this setup, presenting the results for waveguide and free space broadcasts.
In this experiment we first perform the measurements using waveguides. In the waveguide setup, we can eliminate or significantly reduce external factors such as atmospheric attenuation and signal dispersion, allowing us to isolate and study the inherent characteristics of the broadcast. This controlled environment enables us to gain insights into the behavior of the thermal state QKD protocol without the confounding effects of external interference.

Free space transmission presents additional challenges such as losses and interference and additional noise sources, nevertheless we have achieved short range broadcasts. 
It is worth noting that short-range free space transmission is not a limitation, in fact short range QKD would be well-suited for applications like payment systems, car remotes, Wi-Fi, and Bluetooth, where communication typically occurs within a limited distance. In these scenarios, the challenges of atmospheric attenuation and signal dispersion are less pronounced, making short-range free space transmission a practical and efficient choice.



\section{Quadrature Phase Shift Keying}
\begin{figure}[H]
	\centering
	\includegraphics[width=0.3\textwidth]{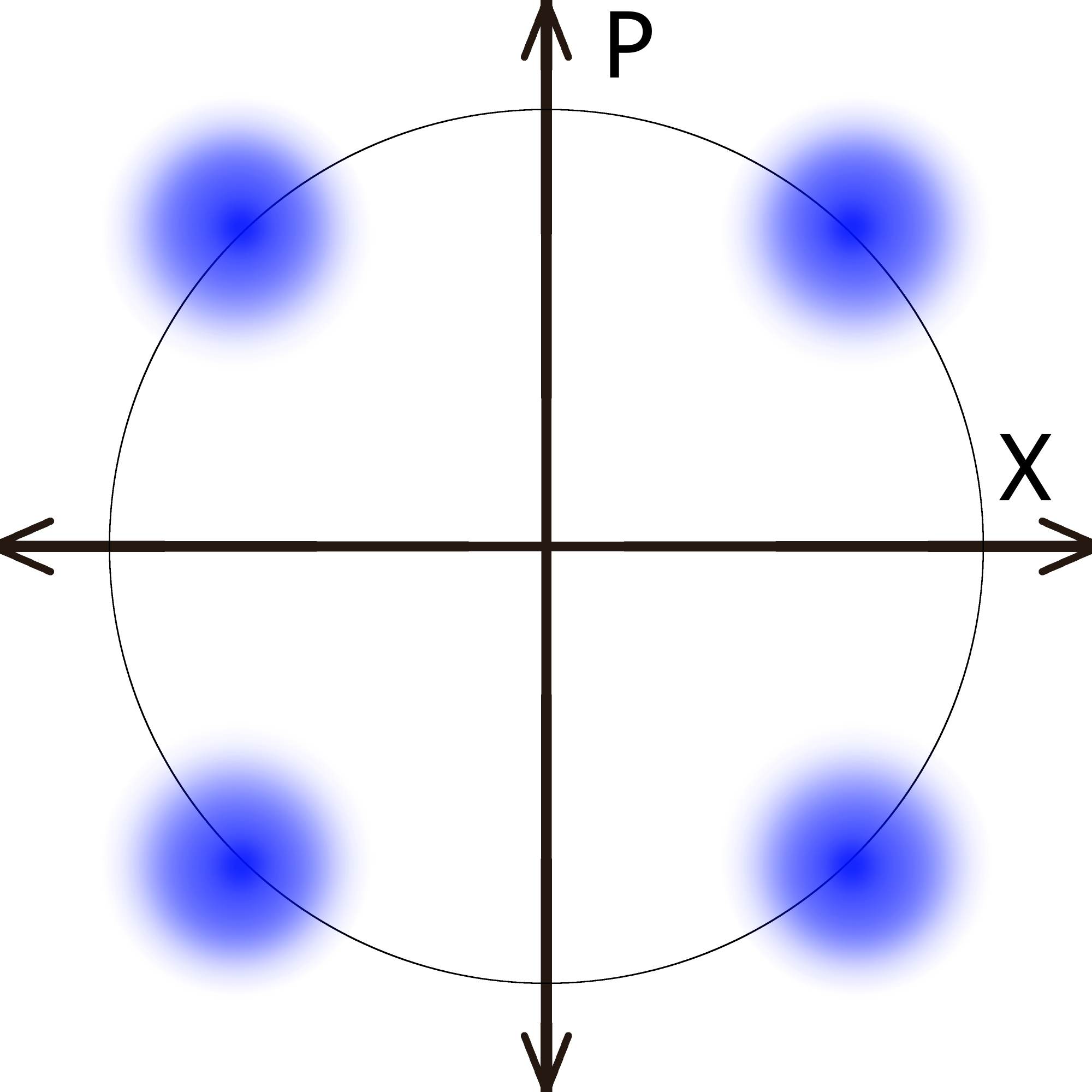}
	\caption{\textbf{QPSK.} A form of PSK, Quadrature Phase-Shift Keying, where two bits of information are sent per signal through assigning one of the four possible combinations of two bits to each of the four clusters. The axes refer to the amplitudes of a pair of sinusoidal waves which differ in phase by $\frac{\pi}{2}$. These amplitudes are adjusted to produce different signals.  \label{fig:QPSK}
	} 
\end{figure}
The method that we will use to encode the information is the classical communication system known as Quadrature Phase Shift Keying (QPSK). 
This is a modulation scheme widely used for transmitting digital data over radio frequencies. 
QPSK represents a set of techniques where data is encoded by varying the phase of a carrier wave among four possible values, typically $0^\circ$, $90^\circ$, $180^\circ$, and $270^\circ$. Each of these phase values corresponds to a different symbol (see Fig. \ref{fig:QPSK}). There are four states, hence this allows for the transmission of two bits per symbol.

Using the language of quantum optics, we can also describe QPSK in terms of displaced thermal states. 
A displaced thermal state arises when a thermal state undergoes displacement by a coherent state. 
A thermal state describes a system in thermal equilibrium with its surroundings, characterized by a distribution of energy levels following a thermal distribution.
In the context of QPSK, we can draw an analogy to displaced thermal states by considering the phase diagram commonly used in quantum optics. In this diagram (Fig. \ref{fig:QPSK}), the horizontal axis represents the real part of the amplitude of the quantum state, while the vertical axis represents the imaginary part. Each point in this phase space corresponds to a unique quantum state.

Now, let's consider the four phase shifts used in QPSK modulation: $0^\circ$, $90^\circ$, $180^\circ$, and $270^\circ$. Each of these phase shifts can be viewed as a displacement of a thermal state by a coherent state with a specific phase difference. For example, a phase shift of 0° corresponds to a normal displacement, and adding a $\pi/2$ phase shift to the carrier corresponds to displacement by a coherent state with a phase difference of $90^\circ$.

\begin{figure}[htb]
	\centering
	\includegraphics[width=0.5\textwidth]{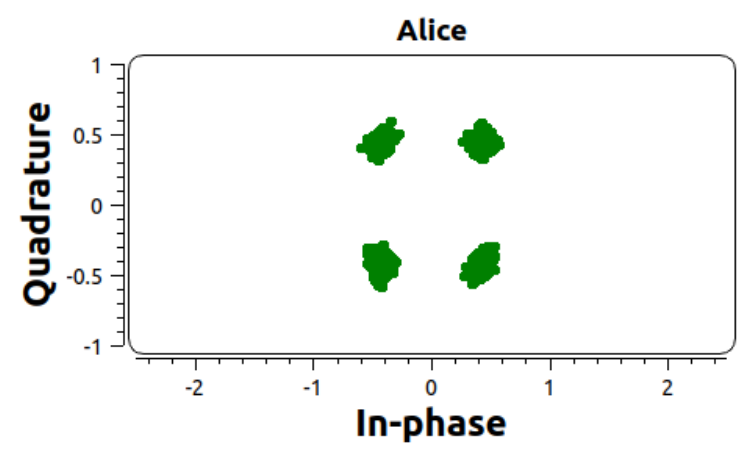}
	\caption{\textbf{Thermal Clusters.} An snapshot of the output of the thermal source after modulation. A constellation modulator produces the four clusters expected in QPSK, as shown in Figure \ref{fig:QPSK}. \label{fig:Clusters}
	} 
\end{figure}

By encoding data using different phase shifts, QPSK effectively manipulates the quantum states in phase space, allowing for the transmission of digital information. 
This type of encoding is the current standard for a digital communications system. 
It therefore becomes relatively straightforward to produce and measure a displaced thermal state with high accuracy and Figure \ref{fig:Clusters} shows a measurement of four displaced states using an off-the-shelf QPSK receiver.

The benefit of QPSK encoding is that binary data can be encoded as `quadratures'. Four quadratures give 2 bits of binary information. This allows us to `error correct' the transmission phase. The transmission of a random binary code leads to a unique unravelling of the phase sequence which therefore allows Alice and Bob to align their data. This is a critical component of thermal state QKD, because they must be able to locate the $\Delta t=0$ peak in the $g_2$ correlation spectrum \cite{Newton2019}. The unique phase unravelling therefore allows them to uniquely locate the peak correlation (see for example Figure \ref{fig:Correlations}).

\begin{figure}[htb]
	\centering
	\includegraphics[width=0.5\textwidth]{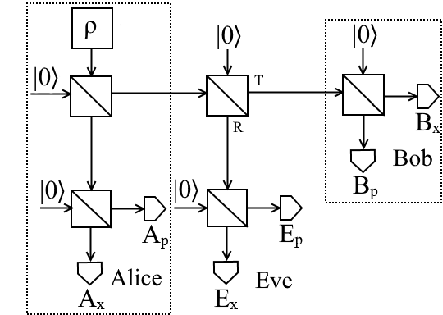}
	\caption{\textbf{Method.} A diagram of the central broadcast thermal protocol. A beam from a thermal source is incident on a beamsplitter, with outputs sent to Alice and Bob. Eve intercepts Bob's beam with a beamsplitter of transmittance T, with the other beamsplitters being 50:50. This diagram was originally used in ”Thermal state quantum key distribution” \cite{Thermal_State_QKD}, and is licensed under CC-BY 4.0. \label{fig:Protocol}
	} 
\end{figure}

Having established the encoding that we are using, the method employed in this study follows a well-established framework for thermal state quantum key distribution (QKD) and its variant using displaced thermal states \cite{Thermal_State_QKD,Displaced_Thermal_State_QKD}. 
A simplified flow chart illustrating the communication process is provided in Figure \ref{fig:Protocol}. 
A displaced thermal source is directed onto a beam splitter, with the resulting output channels connecting to Alice and Bob, who aim to establish secure communication. 
Before Bob can perform any measurements, Eve attempts a beam splitter attack. 
Each party subsequently uses heterodyne detection to obtain a series of correlated measurement pairs, denoted as $\left(x_{i},p_{i}\right)$. These measurements are then processed to derive correlated bit strings by computing $ z_{i} = \sqrt{x_{i}^2+p_{i}^2} $ for each measurement pair. 
We use coarse grained slicing method where a binary value of 0 or 1 is assigned to each $ z_i $ value based on whether it falls above or below the median value obtained at that detector. 

There are two protocols that we could use to transmit continuous quantum information, a `prepare-and-measure protocol' where Alice generates random Gaussian states and transmits them to Bob and the thermal\cite{Newton2019,Newton2020} or passive state QKD protocol\cite{Qi2018,Qi2020}.
From Bob's perspective, the two protocols are identical due to the statistical equivalence between the Gaussian states transmitted by Alice and the passive states used in the passive state QKD protocol.

From Bob's point of view, in both scenarios, he receives quantum states that exhibit Gaussian statistics. These states are characterized by their mean and variance, which encapsulate information about the transmitted quantum information. 
In both cases, Bob's task is to perform measurements on the received quantum states to extract the relevant information for key generation. While in the prepare-and-measure protocol, 
Bob measures the received Gaussian states using compatible measurement bases, typically chosen randomly. He then records the measurement outcomes and communicates with Alice to establish a shared secret key through classical post-processing techniques.

Similarly, in the passive state QKD protocol, Bob receives passive states generated by Alice, which also exhibit Gaussian statistics. Bob performs measurements on these received states and follows the same procedure as in the prepare-and-measure protocol to extract the shared secret key.
Overall, despite differences in the physical implementation of the two protocols (active state generation in prepare-and-measure versus passive state generation in passive state QKD), from Bob's perspective, the statistical properties of the received quantum states are identical. 
This equivalence allows Bob to employ the same measurement and key generation procedures in both scenarios, resulting in similar operational outcomes for the two protocols. With this in mind there are security proofs over a wide range of performance metrics which support the concept of a thermal state resource\cite{Zhong2021,Qi2018,Qi2020,Xu2021,Bai2019,Wu2020,Wu2021,Ghesquiere2021}.

To perform the experiment, we use USRP-2901 radio transceivers broadcasting at a frequency of 2GHz using a PRS10 Rubidium Oscillator as a time reference and GNU Radio is used for signal processing (the signal processing flowchart is shown in Appendix \ref{fig:Chart}). 
We use Costas loops and polyphase clock sync blocks to stabilise the phase and synchronise the measurements. 

The first experimental results were obtained using waveguides between Alice, Bob, and Eve. 
This has two primary effects. The first is that we can be sure that all of the signal that does not go to Bob, goes to Eve (this is not possible in a free space apparatus) and it therefore conforms with the standard requirements for testing the security of CVQKD \cite{Grosshans2002,Usenko2015,Grosshans2003}. Secondly it provides a very low phase noise environment to establish the operational parameters. 

The two main limitations with the waveguide channel are: firstly the broadcast signal is attenuated to avoid sending large signals into the receivers (maximum input signal is -30dBm, or $1\mu W$) and secondly thermal variations during the measurement lead to phase and amplitude fluctuations as a result of small changes in the length of the waveguides. The effect of this can be seen as a phase `hopping' in Figure \ref{fig:Correlations}.

The free space channel uses an omni-directional antenna at both the transmitters and the receivers side, this means that there is no specific directionality to the signal and allowing Bob to be located anywhere around Alice. This means that neither Eve nor Bob has a specific advantage. However this comes at the expense of substantial signal losses limiting the range of the transmission.

\section{Experimental Results and Security Analysis} \label{Results}

The constellation modulator produces four displaced states which are equally spaced around a circle centred on the origin in phase space. 
A snapshot of the received QPSK broadcast is shown in Figure \ref{fig:Clusters}. 
We processed quadrature measurement results by rotating the angles of the four QPSK elements so that they overlap, effectively creating a single displaced state. 

However before we could do this we needed to establish a time synchronisation between sender and receiver. A 2 GHz signal has a wavelength of 15cm, hence there can be several oscillations of $2\pi$ between transmission and reception, essentially randomising the phase relationship between Alice and Bob. 
Therefore, in order to synchronise the measurement times, Alice and Bob compare the digital signals obtained by observing the quadrant phase. 
The string of measurement results that were obtained presented a unique signature of the timing and this allowed Alice and Bob to correct for time delays. Revealing the alignment of the phase quadratures to the eavesdropper does not affect security as the secret key will be derived using correlated signal noise.

Once the time delay is accounted for, bit strings are derived from the correlated amplitude measurements. From these bit strings, Shannon mutual information, $I(X;Y)$, are calculated which are used to test if the protocol is successful. For a key to be distilled from the bit strings after employing advantage distillation, we require that the conditional mutual information, $I\left(A;B|E\right)$ is greater than zero \cite{Maurer99}.

\begin{figure}[H]
	\centering
	\includegraphics[width=0.8\textwidth]{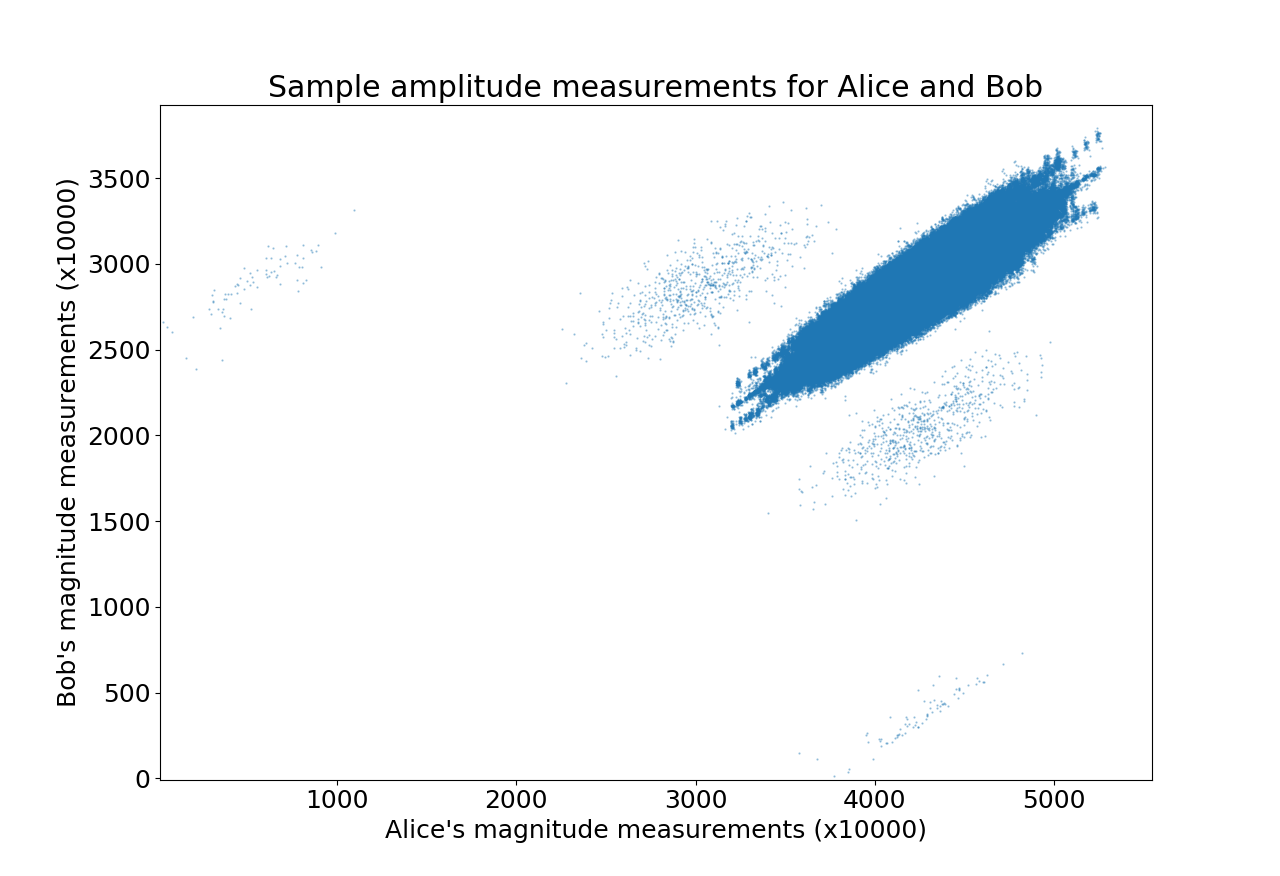}
	\caption{\textbf{Correlations in thermal states.} A comparison of a sample of amplitude measurements performed by Alice and Bob after adjusting for time errors, displaying Hanbury Brown and Twiss correlations. The data displays a number of correlated overlapping features each with a high level of correlation. Small phase hops occur between the source and the Bob-Eve beam splitter probably due to thermal changes in the waveguide, giving rise to random phase drifts. For short times the phase remains stable and this can be seen as drifting correlation. In addition to this are stray reflections in Alice's transmission line which create faint ghost images that can be seen either side of the main peak. These have very little effect on the overall correlation, because the ghost reflections represent only a tiny fraction of the data. \label{fig:Correlations}
	} 
\end{figure}

\begin{figure}[H]
	\centering
	\includegraphics[width=0.8\textwidth]{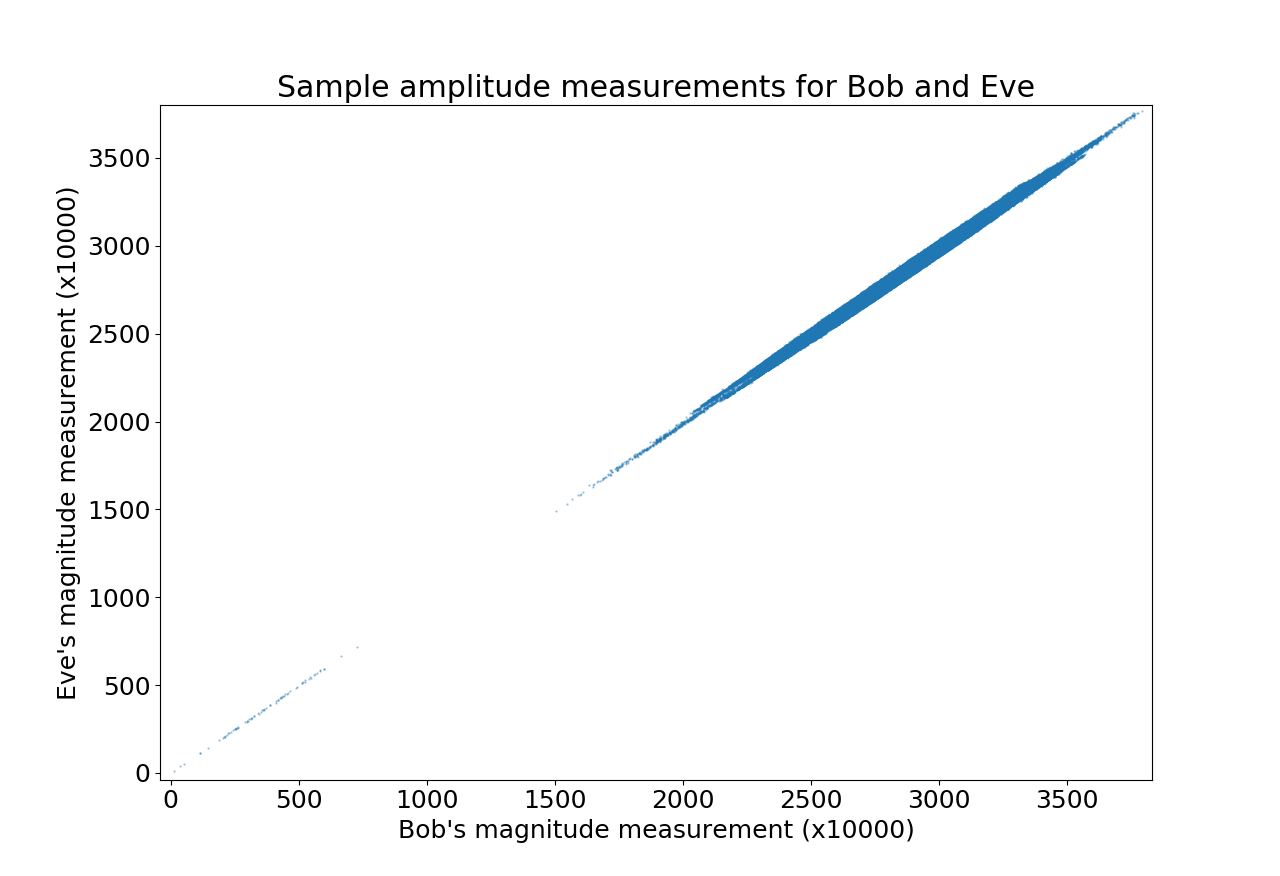}
	\caption{\textbf{Waveguide measurements for Bob and Eve.} A comparison of a sample of amplitude measurements performed by Bob and Eve after adjusting for time errors. As there is no practical difference between Eve and Bob therefore there are fewer amplitude errors and the long duration signal is highly correlated due to the higher degree of symmetry between their respective detectors.   \label{fig:BobEveWired}
	} 
\end{figure}

For continuous data it is possible to calculate the $g_2$ correlation \cite{Newton2019}, however this data is discrete making the $g_2$ inaccessible. Figure \ref{fig:Correlations} shows a plot comparing Alice and Bob's amplitude measurements in a sample set of data ($n=3\times10^6$ points) after compensating for time delay errors. These measurements are highly correlated (r= 0.9264), clearly displaying the correlations expected for the Hanbury Brown and Twiss effect, with Bob and Eve's measurements showing the same behaviour (r=0.99362), as shown in Figure \ref{fig:BobEveWired}. For the data presented here, $I\left(A;B|E\right)=0.04688$ and $\Delta I=I\left(A;B\right)-I\left(B;E\right)=0.18154$. While this specific example meets the standard QKD requirements for key exchange\cite{Grosshans2002}
\cite{Usenko2015}
\cite{Grosshans2003}, shot to shot variability in the measurements resulting from thermal drift in the waveguide (see fig. \ref{fig:Correlations}) means that this is not always the case as $\Delta I$ occasionally fluctuates into negative numbers. However, the combination of advantage distillation and privacy amplification ensures Alice and Bob are able to retrieve a key in general. Therefore given that the waveguide tests showed some inconsistencies we transitioned to a freespace broadcast channel for more reliable results, which in any case, is the more realistic scenario for secure key exchange.

The experimental setup was adjusted so that the output from Alice's thermal source is now connected to a whip antenna which broadcasts an omni-directional signal that can be detected by Bob and Eve. Alice still uses a waveguide to her local receiver in this model, in agreement with the assumption that Alice is in control of the source. 

A whip antenna typically exhibits an omnidirectional emission pattern in the horizontal plane meaning that it radiates electromagnetic waves uniformly in all directions around the antenna axis, hence there is no longer a preferred direction and Bob and Eve are free to move around the source. 
The downside of this is that the cross section of Bob's antenna is relatively small and therefore he experiences much higher loss.
However, as we are comparing physical measurements made by the eavesdropper to those made by Bob, the free space setup makes security less challenging owing to the reduction in the correlations between Bob and Eve's measurements. To some extent this situation lies outside of Eve's control.
Here we are following a ``realistic'' eavesdropper scenario\cite{Ghalaii2023}, noting that it is unreasonably obvious for the eavesdropper to try to collect more of the signal than Bob over a small distance.

\begin{figure}[H]
	\centering
	\includegraphics[width=0.4\textwidth]{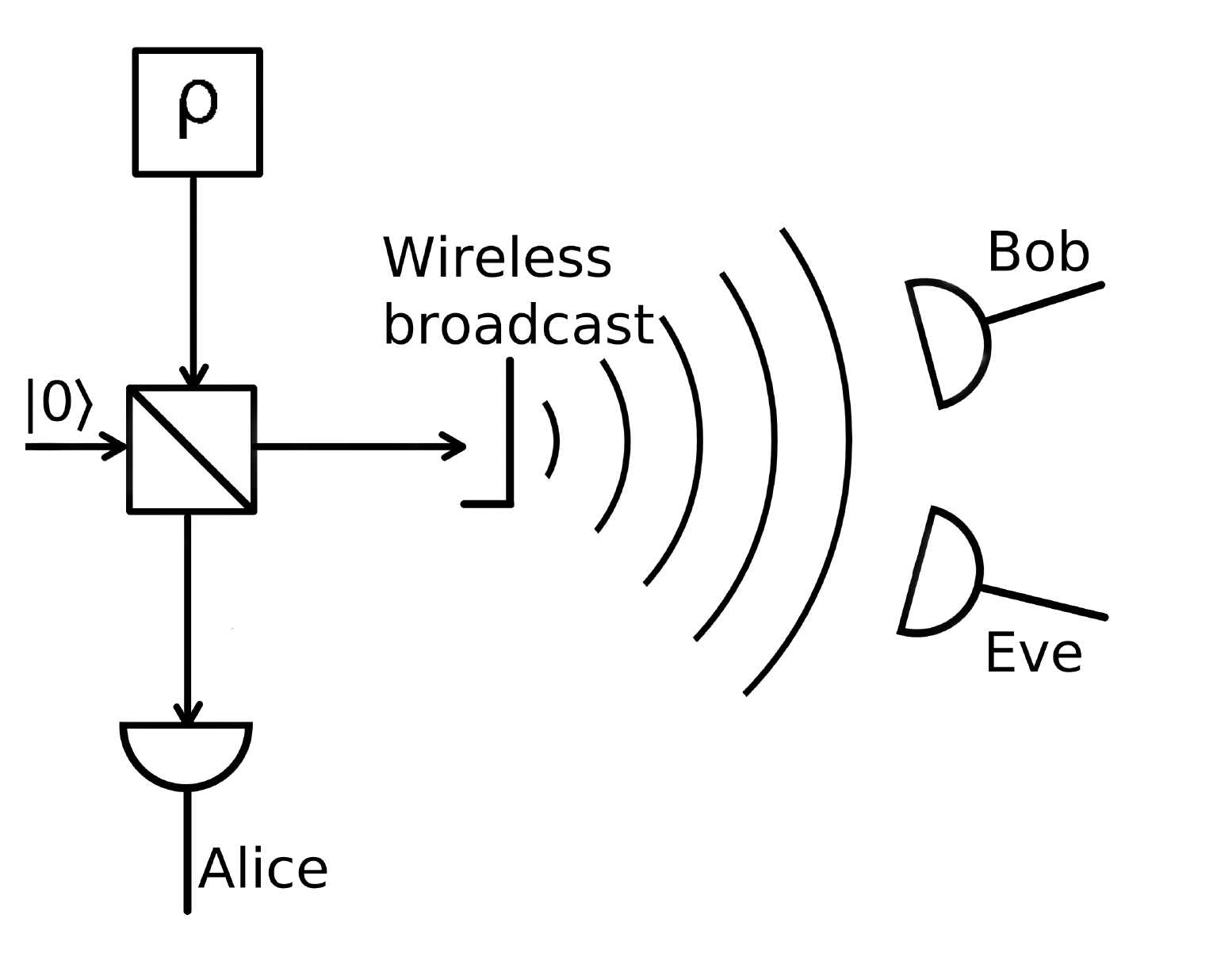}
	\caption{\textbf{The Free space apparatus:} The adjusted version of the apparatus, in which the waveguide channel to Bob is replaced with an antenna. Two antennas connected to a second USRP represent Bob and Eve. \label{fig:WirelessProtocol}
	} 
\end{figure}

We performed a free space broadcast over a distance of 1 metre at a sample rate of 250k samples per second. The limited range was a result of the amount of reflective surfaces in the lab, the source power and antenna configurations. As noted above, this is still a practical range for several applications. Moreover, the number of reflective surfaces reduce substantially when the device is used `outdoors' allowing a greater range in practise. As with the previous setup, no error correction is employed beyond compensating for time delay and the phase shifts added during the protocol. 

\begin{figure}[H]
	\centering
	\includegraphics[width=0.8\textwidth]{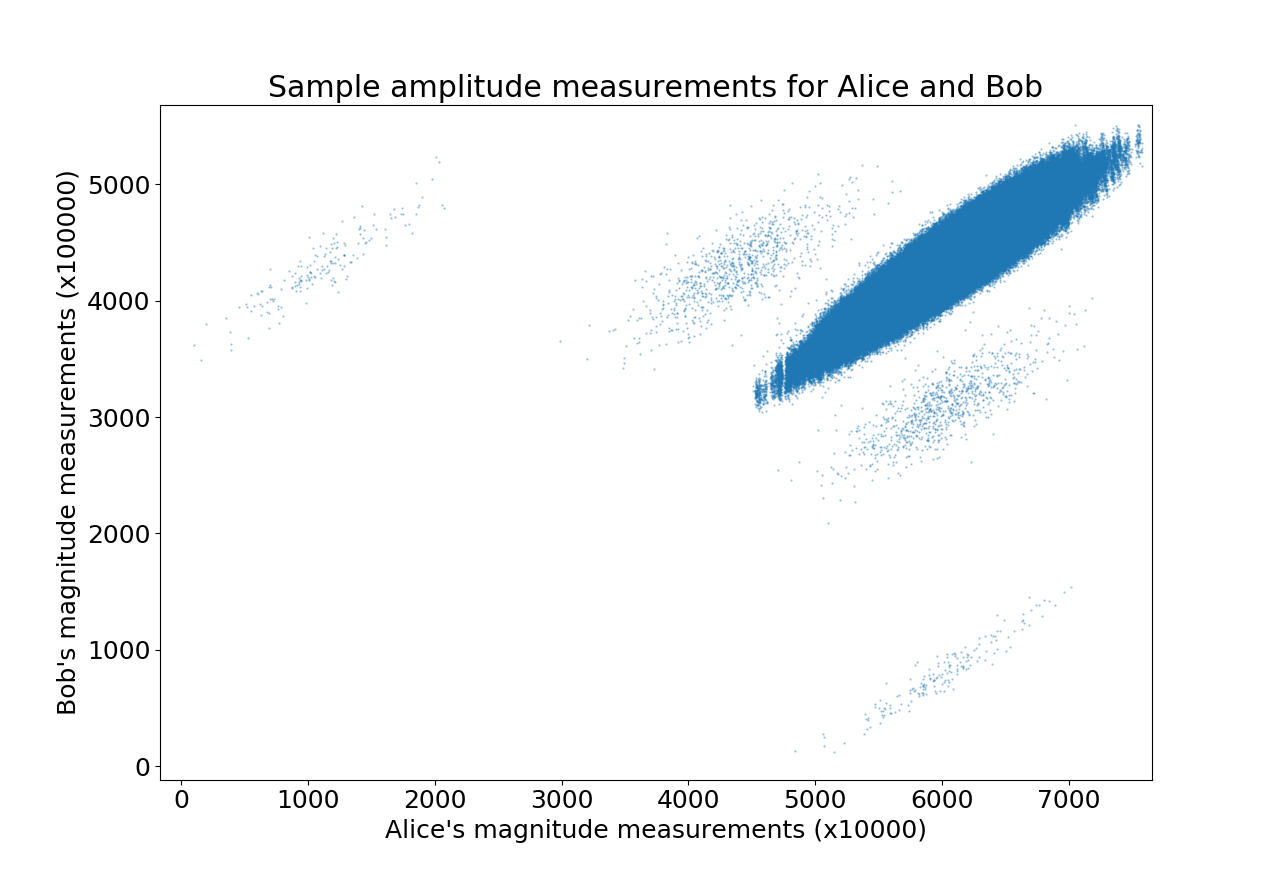}
	\caption{\textbf{Correlations in thermal states.} A sample (n=3000000) of amplitude measurements performed by Alice and Bob. While the measurement results were less correlated than the waveguide version displayed in Figure \ref{fig:Correlations}, they were still suitable for key distribution. The amplitude measurement results are shown in Figure \ref{fig:Histogram}. Again ghost reflections can be seen in the data similar to the waveguide model and therefore are most likely a result of an impedance mismatch in Alice's apparatus.  \label{fig:Wireless}
	} 
\end{figure}

\begin{figure}[H]
	\centering
	\includegraphics[width=0.8\textwidth]{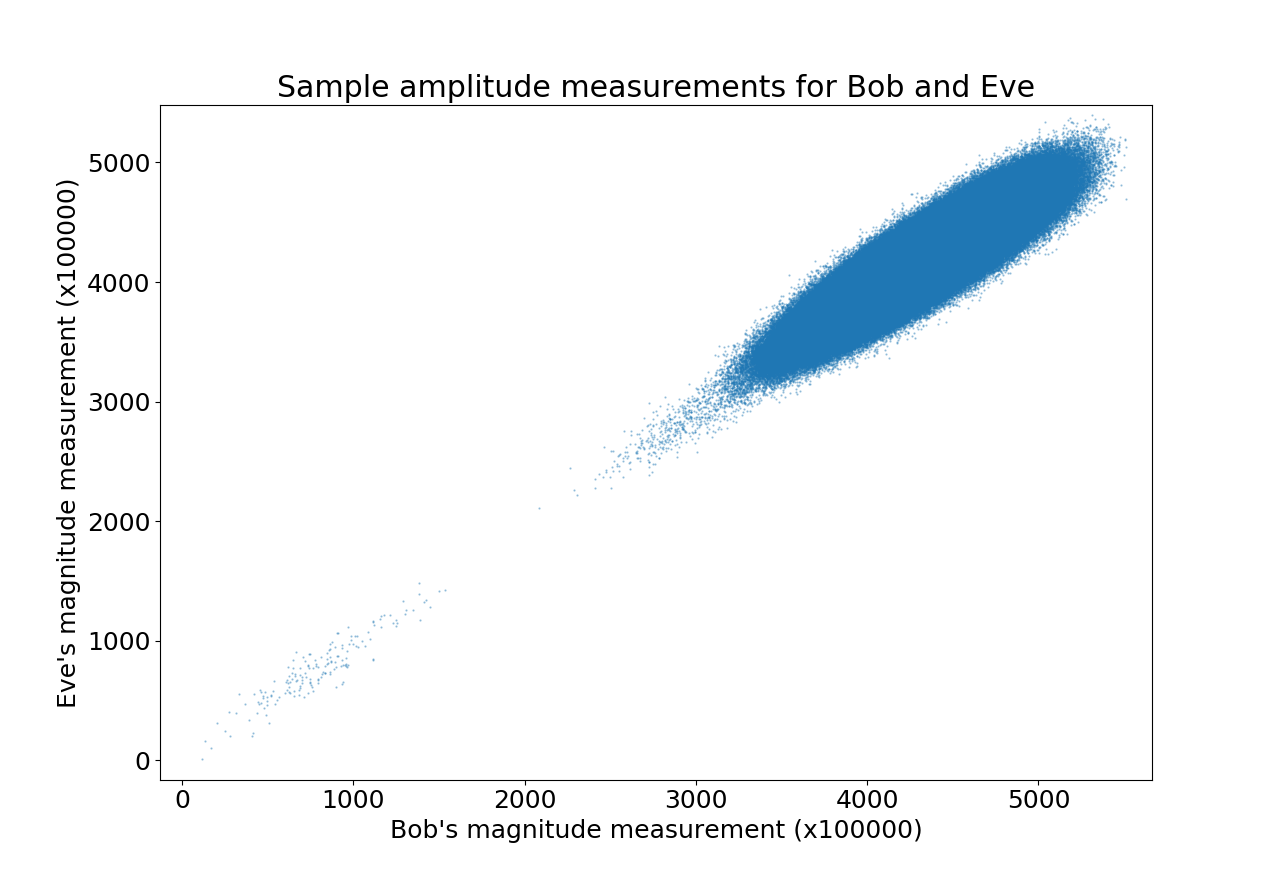}
	\caption{\textbf{free space measurements for Bob and Eve.} A sample (n=3000000) of amplitude measurements performed by Bob and Eve. The additional loss results in less correlated measurements. \label{fig:BobEveWireless}
	} 
\end{figure}

While outcomes in general displayed higher variance than in the waveguide version, bit strings produced through this method were still suitable for conversion into keys with reverse reconciliation. Comparing Alice and Bob's measurements shows very similar behaviour in both the waveguide and free space setups, as seen in Figure \ref{fig:Wireless}. However, a major impact of the change to free space was the decrease in correlation between Bob and Eve, dropping from near-identical bit strings down to a mean correlation coefficient of $r=0.89$. A comparison of a sample of such measurements are shown in Figure \ref{fig:BobEveWireless}.

Additionally, we see faint copies of the measurement cluster repeated at several other points around the plot. This is likely due to multipath propagation, in which the receiver detects copies of the broadcast signal that have been reflected by other objects and is a source of error which is difficult to completely remove. This is a well known problem in free space communication and its detection here is not surprising. There are multiple ways to reduce the impact of this, such as through repeated hopping of broadcast frequencies \cite{Multi}, or in more controlled circumstances, RF absorbing material can be used to reduce reflections. While this effect does negatively impact the performance of a free space broadcast, it should be noted that Eve would be similarly affected due to also having no control over such reflections.

The free space variant was repeated over a distance of 1 meter, producing 20 sets of bit strings with approximate length of $3\times10^6$ bits each (12 seconds of data). From this, we calculated mutual information values and correlation coefficients, giving a mean conditional mutual information $I\left(A;B|E\right)$ of $0.126\pm 0.046$. As this is a positive value, we have achieved a sufficient condition for secure communication \cite{Maurer2007}.

For direct reconciliation we found $I\left(A;B\right)-I\left(A;E\right)\approx 0$. This is expected due to the symmetry in the system. For reverse reconciliation, we found $I\left(A;B\right)-I\left(B;E\right)=0.082\pm 0.06$. Meeting the conditions for secret key exchange. We found a bit error rate of approximately $11.3\%\pm 2.9\%$) between Alice and Bob.

Afterwards, the free space transmission range was extended to 5 meters. We found that as the distance increased the timing errors increased and correction became significantly more difficult, therefore maintaining accurate phase and timing corrections became challenging. 
Despite this, we still managed to generate suitable bit strings. Beyond 5 meters we would need to implement a more sophisticated error correction mechanism to address synchronization issues between Alice's and Bob's measurements. Such corrections are routine in digital communications (mobile phones would not work without it) but such integration will have to wait for a future iteration of the communication protocol.

\section{Conclusions} \label{Conclusions}
In summary, we have demonstrated the use of displaced thermal states as a resource for quantum key distribution in a microwave field. By harnessing thermal sources in a `Passive State QKD' configuration, we present a simplified and feasible approach for secure communication, particularly in scenarios requiring short-range applications which demonstrated the practical implementation of thermal QKD. 

An interesting outcome is that this potentially enables the direct integration of QKD into existing communication systems, rendering it accessible and viable for real-world deployment. The QKD component and digital communications components are (almost) independent of one another with digital error correction increasing the alignment of the states and therefore improving the secret key rate. 

Looking ahead, avenues for exploration include extending the range of free space QKD and leveraging error-correcting codes to enhance performance.
Our implementation of QKD within an off-the-shelf digital communications exploits the analogy between QPSK modulation and displaced thermal states, creating a relatively seamless integration between quantum and classical communication protocols. This interplay provides valuable insights for future developments in the field and paves the way for further innovation.

\section*{Acknowledgements}
This work was supported by the Northern Triangle Initiative Connecting capability fund as well as funding from the UK Quantum Technology Hub for Quantum Communications Technologies EP/M013472. The data used to plot the graphs in Figures \ref{fig:Correlations}, \ref{fig:BobEveWired}, \ref{fig:Wireless}, and \ref{fig:BobEveWireless} are available on request from the author.

\newcommand{\newblock}{}
\bibliographystyle{unsrtnat}
\bibliography{References}

\appendix
\section{Signal Processing}

\begin{figure}[H]
	\centering
	\includegraphics[width=1.0\textwidth]{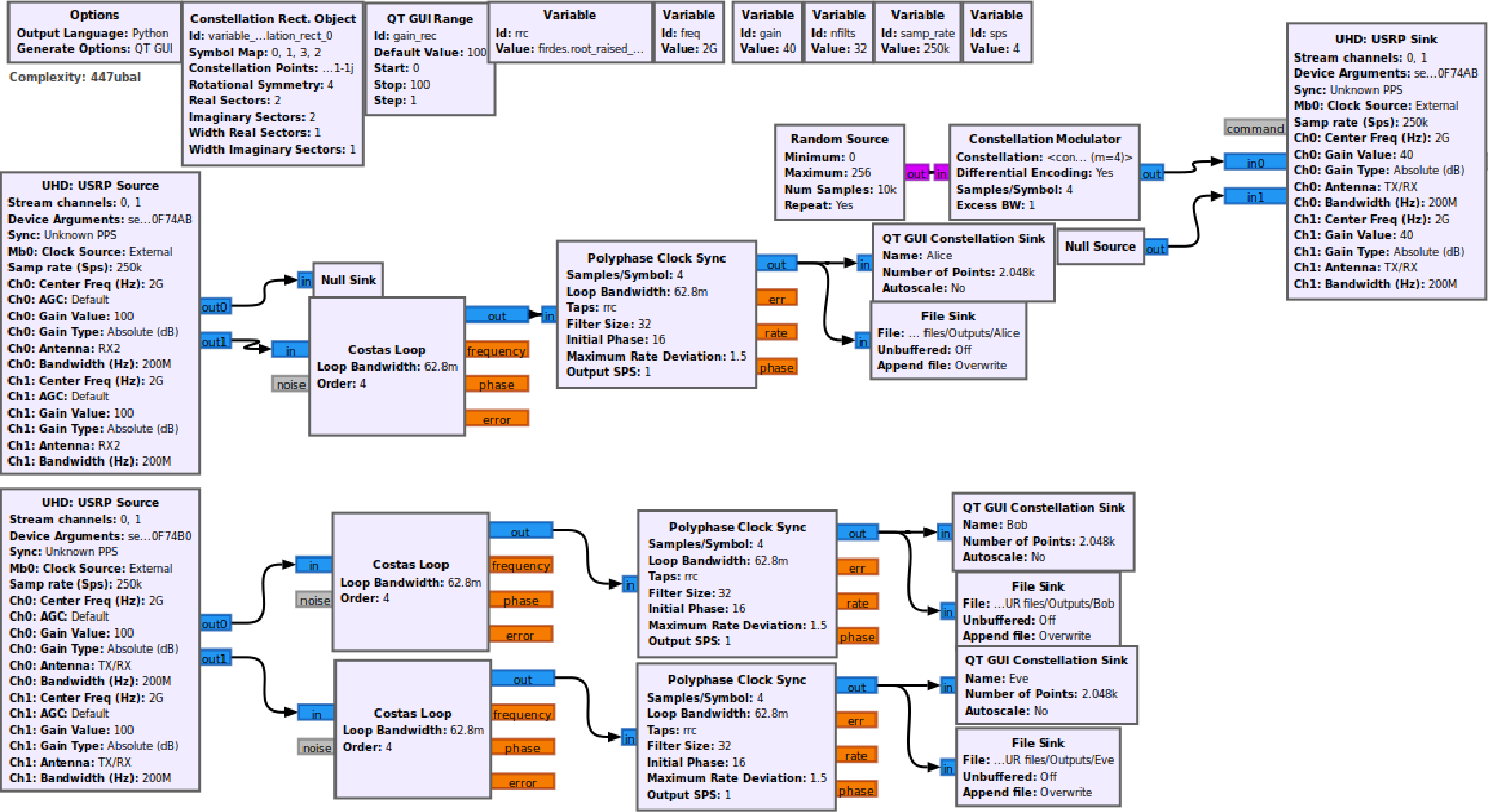}
	\caption{\textbf{The GNU Radio flowchart.} Signal processing performed by GNU Radio. Four clusters are sent to Alice, Bob and Eve for measurement. \label{fig:Chart}
	} 
\end{figure}

\section{free space Histogram}

\begin{figure}[H]
	\centering
	\includegraphics[width=1.0\textwidth]{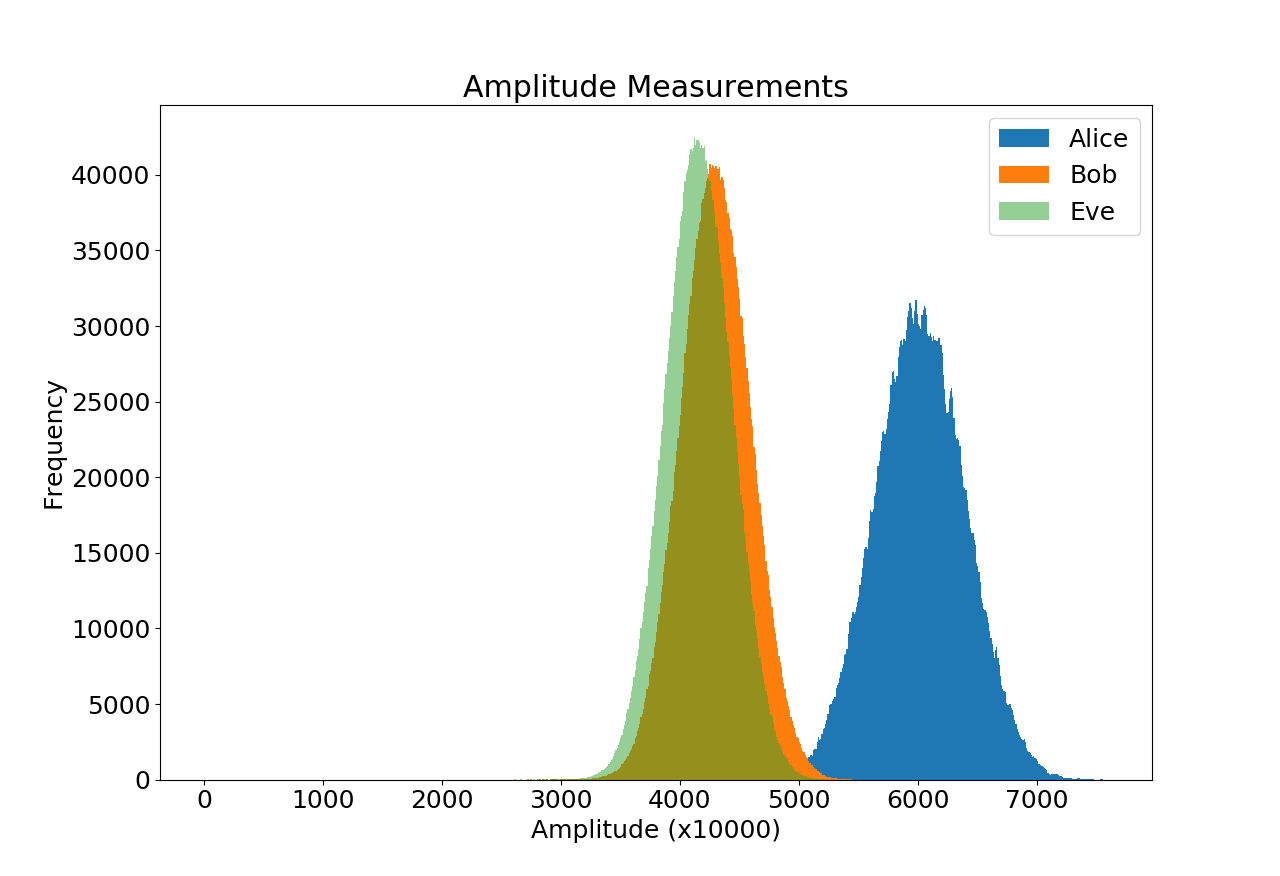}
	\caption{\textbf{free space thermal states broadcasting.} A sample (n=3000000) of recorded amplitude measurements. .
		\label{fig:Histogram}
	} 
\end{figure}

\end{document}